# Computation of dynamical structure factors with the recursion method

A Fledderjohann, M Karbach, K-H Mütter † and P Wielath

Physics Department, University of Wuppertal

42097 Wuppertal, Germany

**Abstract.** We compute the energies and transition probabilities for low excitations in the one dimensional antiferromagnetic spin-1/2 Heisenberg model by means of the recursion method. We analyse finite size effects in the euclidian time ($\tau$)-representation and compare the resulting estimate for the thermodynamical limit with two parametrizations for the dynamical structure factors in the spectral ($\omega$)-representation.

Short title: Computation of dynamical structure factors

July 11, 1995

† muetter@wpts0.physik.uni-wuppertal.de

1. Introduction

Quantum spin systems with known dynamical behaviour are rare. The spin 1/2 XX-model can be mapped on a free fermion system [1] and the dynamical spin-spin correlators can be computed analytically [2-5]. A second example is the Haldane-Shastry model [6,7] -- an isotropic Heisenberg model with couplings which decrease with the inverse square of the distance between two spin operators. In this model only two-spinon excited states contribute to the zero temperature dynamical structure factor [8,9]. In this paper we are concerned with the familiar antiferromagnetic Heisenberg model with nearest neighbour coupling:

$$H = 2 \sum_{x=1}^{N} \vec{s}(x)\vec{s}(x+1) \tag{1.1}$$

and periodic boundary conditions. The dynamical structure factors at $T = 0$:

$$S(\omega, p, N) = \sum_n \delta(\omega - (E_n - E_0))|\langle n|s_3(p)|0\rangle|^2 \tag{1.2}$$

are defined by the transition probabilities $|\langle n|s_3(p)|0\rangle|^2$ from the groundstate $|0\rangle$ to the excited state $|n\rangle$ with energies $E_0$ and $E_n$, respectively. The transition operator

$$s_3(p) = \frac{1}{\sqrt{N}} \sum_{x=1}^{N} \exp(ipx) s_3(x) \tag{1.3}$$

is just the Fouriertransform of the spin-operator $s_3(x)$ at site $x$. The lower bound for the excitation energies

$$\omega_1 = \omega_1(p) \equiv \pi \sin p \tag{1.4}$$

has been computed by Cloizeaux and Pearson [10]. The infrared behaviour at $p = \pi$:

$$S(\omega, p = \pi, N = \infty) \overset{\omega \to 0}{\leadsto} \omega^{-1} \tag{1.5}$$

has been investigated in reference [11]. G. Müller and collaborators [12] proposed an ansatz, which takes into account only two spinon excitations:

$$S(\omega, p) = A \frac{\Theta(\omega - \omega_1)\Theta(\omega_2 - \omega)}{\sqrt{\omega^2 - \omega_1^2}}. \tag{1.6}$$

The upper bound of these excitations

$$\omega_2(p) \equiv 2\pi \sin \frac{p}{2} \tag{1.7}$$

has been obtained from the Bethe ansatz. The ansatz (1.6) has been shown [13] to satisfy certain conditions on the spectral moments for the dynamical structure factor. Moreover, it was successfully applied to the description of neutron scattering data [14]. On finite systems, the dynamical structure factors are defined as $\delta$-function contributions in $\omega$. In order to extract the thermodynamical limit it is useful [15] to consider the Laplace transform of (1.2):

$$S(\tau, p, N) = \int_{\omega_1}^{\infty} d\omega \; S(\omega, p, N) \exp(-\omega\tau) \qquad (1.8)$$
$$= \langle 0 | s_3^+(p) \exp[-\tau(H - E_0)] s_3(p) | 0 \rangle.$$

It can be interpreted as a euclidean time ($\tau$)-representation of the dynamical structure factor.

In this paper we want to propose a new method to compute the dynamical structure factors. It is based on the recursion used in [16] as input for the continued fraction approach. In section 2 we will demonstrate, how the energies and transition probabilities for the low excitations can be obtained directly from the recursion approach. In section 3 we will present our results on the dynamical structure factors (1.8) for systems with $N = 4, 6, 8, ..., 28$. Finite size effects will be analyzed and an estimate for the thermodynamical limit will be given. This estimate is compared in section 4 with the prediction (1.6) of Müller *et al.*. We also propose a modification of this ansatz, which yields better agreement with the estimate of the dynamical structure factors in the thermodynamical limit.

## 2. The Recursion method

Following [16], (1.8) can be computed by iteration. For this purpose one expands the euclidean time evolution of the 'initial' state $|f_0\rangle \equiv s_3(p)|0\rangle$:

$$\exp[-\tau(H - E_0)] s_3(p) |0\rangle = \sum_{k=0}^{\infty} D_k(\tau) |f_k\rangle \qquad (2.1)$$

in terms of an orthogonal basis $|f_k\rangle$ which is constructed recursively by application of the Hamilton operator:

$$(H - E_0)|f_k\rangle = |f_{k+1}\rangle + a_k |f_k\rangle + b_k^2 |f_{k-1}\rangle, \quad k = 0, 1, 2, ... \qquad (2.2)$$

$$a_k = \frac{\langle f_k | H - E_0 | f_k \rangle}{\langle f_k | f_k \rangle}, \quad k = 0, 1, 2, ... \qquad (2.3)$$

$$b_k^2 = \frac{\langle f_k | f_k \rangle}{\langle f_{k-1} | f_{k-1} \rangle} \quad k = 1, 2, ..., \qquad (2.4)$$

with $|f_{-1}\rangle \equiv 0$ and $b_0^2 \equiv 0$. On finite systems the iteration (2.2) will terminate after $L$ steps, where $L$ is the dimension of the Hilbert space spanned by the states $(H - E_0)^l s_3(p)|0\rangle$, $l = 0, 1, ..., L-1$. The iteration generates a tridiagonal $L \times L$-matrix $O$:

$$(H - E_0)|f\rangle = O|f\rangle \qquad (2.5)$$

with matrix elements:

$$\langle f_{k'}|O|f_k\rangle = \begin{cases} 1: & k' = k+1 \\ a_k: & k' = k \\ b_k^2: & k' = k-1 \\ 0: & \text{else} \end{cases} \qquad (2.6)$$

in the basis $|f_k\rangle$. The eigenvalues of $O$ yield the excitation energies $\omega = E_n - E_0$ and the eigenvectors represent the excited states $|n\rangle$ of $(H - E_0)$ in the basis $|f_k\rangle$:

$$|n\rangle = \sum_{k=0}^{L} |f_k\rangle \frac{\langle f_k|n\rangle}{\langle f_k|f_k\rangle}. \qquad (2.7)$$

The zero components $\langle f_0|n\rangle$ of the eigenvectors and the excitation energies $E_n - E_0$ determine the dynamical structure factor:

$$S(\tau, p, N) = \sum_n |\langle f_0|n\rangle|^2 \exp[-\tau(E_n - E_0)]. \qquad (2.8)$$

Performing the iteration (2.2) numerically, one finds that the orthogonality of the vectors $|f_k\rangle$ is lost after a certain number of steps due to rounding errors. On the other hand it turns out that the energies $E_n - E_0$ and transition probabilities of the low-lying excitations with $n < 10$ can be obtained already with high accuracy by truncating the problem. We stop the iteration after $\tilde{L}$ steps and diagonalize the truncated $\tilde{L} \times \tilde{L}$ matrix $O$. As an example we show in table 1 the first 9 excitation energies and normalized transition probabilities

$$w_n(p, N) = \frac{|\langle n|s_3(p)|0\rangle|^2}{S(0, p, N)} \qquad (2.9)$$

on a ring with $N = 16$ sites and momenta $p/\pi = \frac{1}{4}, \frac{1}{2}, \frac{3}{4}, 1$.

The approximate solution given in table 1 is obtaind by truncating the iteration after $\tilde{L} = 40$ steps. We only list in table 1 those digits which agree with the numerically exact solution given in [15]. In other words, the number of quoted digits measures the accuracy of the approximation. Note that the energies and transition probabilities of the first four excitations are reproduced correctly with 8 and more digits. The accuracy for the remaining 5 excitations is less impressive. However, this inaccuracy has practically no effect on our evaluation of the dynamical structure factors in the euclidean time representation (1.8), since the contributions of the higher excitations are suppressed twofold. In addition to the exponential damping factor $\exp(-\omega\tau)$ in (1.8) it turns out, that the transition probabilities themselves drop rapidly with $\omega$. Therefore we expect, that the determination of the excitation energies and transition probabilities by means of the recursion method will yield as well reliable results for larger systems with $N = 18, 20, ..., 28$. In these systems we have checked the energy and the transition probability of the first excitation with momentum $p = \pi$. The result obtained with the recursion method with $\tilde{L} = 40$ iterations is given in table 2. Again we only list those digits which agree with the numerically exact result. The latter is obtained from a determination of the groundstate in the channels with total spin $S = 0$ and $S = 1$, respectively.

**Table 1.** Comparison of leading excitation energies and transition probabilities with the results of complete diagonalization [15] on a ring with 16 sites.

| $\omega_n(\pi/4)$ | $w_n(\pi/4)$ | | $\omega_n(\pi/2)$ | $w_n(\pi/2)$ | |
|---|---|---|---|---|---|
| \multicolumn{3}{c}{$S(\tau=0, p=\pi/4) = 2.982766323178 \cdot 10^{-1}$} | \multicolumn{3}{c}{$S(\tau=0, p=\pi/2) = 6.794375761266 \cdot 10^{-1}$} |
| 2.302618995384 | 9.8645878546337 | $\cdot 10^{-1}$ | 3.3806613858893 | 8.603551483282 | $\cdot 10^{-1}$ |
| 4.29305867857 | 1.0940110661 | $\cdot 10^{-3}$ | 4.197135363571 | 1.288323108707 | $\cdot 10^{-1}$ |
| 4.55019377043 | 8.579155788 | $\cdot 10^{-3}$ | 4.59757074255 | 6.12984208 | $\cdot 10^{-3}$ |
| 5.00373864011 | 7.2716867 | $\cdot 10^{-4}$ | 4.73871528573 | 1.48596359 | $\cdot 10^{-3}$ |
| 5.568925 | 2.792 | $\cdot 10^{-4}$ | 4.902809151 | 9.727121 | $\cdot 10^{-5}$ |
| 5.781 | 4.38 | $\cdot 10^{-4}$ | 5.35132697 | 1.723352 | $\cdot 10^{-4}$ |
| 5.8450 | 1.66 | $\cdot 10^{-3}$ | 5.8903 | 3.01 | $\cdot 10^{-4}$ |
| 6. | | $\mathcal{O}(10^{-4})$ | 5.9 | 1.875 | $\cdot 10^{-3}$ |
| 6. | | $\mathcal{O}(10^{-4})$ | 6.47 | 8. | $\cdot 10^{-5}$ |

| $\omega_n(3\pi/4)$ | $w_n(3\pi/4)$ | | $\omega_n(\pi)$ | $w_n(\pi)$ | |
|---|---|---|---|---|---|
| \multicolumn{3}{c}{$S(\tau=0, p=3\pi/4) = 1.3230534343023$} | \multicolumn{3}{c}{$S(\tau=0, p=\pi) = 4.292303508279$} |
| 2.6381304345681 | 7.717163620986 | $\cdot 10^{-1}$ | 0.540379364500 | 8.01345217378 | $\cdot 10^{-1}$ |
| 3.411532042825 | 7.7506497226 | $\cdot 10^{-4}$ | 2.79206117219 | 1.4109158151475 | $\cdot 10^{-1}$ |
| 4.330655742114 | 1.8574615465811 | $\cdot 10^{-1}$ | 4.668596605332 | 4.6691968837931 | $\cdot 10^{-2}$ |
| 5.03583236201 | 8.985257272 | $\cdot 10^{-4}$ | 5.475947091450 | 1.636197084 | $\cdot 10^{-4}$ |
| 5.440678 | 4.4375 | $\cdot 10^{-4}$ | 5.9070165813 | 1.044901942 | $\cdot 10^{-2}$ |
| 5.46270297 | 3.950172 | $\cdot 10^{-2}$ | 5.994081 | 1.03778 | $\cdot 10^{-6}$ |
| 5.980403 | 8.0786 | $\cdot 10^{-5}$ | 6.573253 | 1.55186 | $\cdot 10^{-4}$ |
| 6.44 | 2.3 | $\cdot 10^{-5}$ | 6.80283 | 6.702 | $\cdot 10^{-5}$ |
| 6.643 | 5. | $\cdot 10^{-4}$ | 7.1 | | $\mathcal{O}(10^{-7})$ |

## 3. Finite size analysis of the dynamical structure factors in the euclidean time representation

For noncritical momenta $p < \pi$ and euclidean times $\tau$ not too large finite size effects are small in the dynamical structure factors (1.8). In figure 1 we show the momentum

Table 2. Comparison of leading excitation energies and transition probabilities with numerically exact results.

| N  | $\omega_1(\pi)$   | $w_1(\pi)$       |
|----|-------------------|------------------|
| 18 | 0.4824988997812   | 0.785322982816   |
| 20 | 0.43589108737     | 0.771429257660   |
| 22 | 0.3975468223202   | 0.7592200871420  |
| 24 | 0.365442071611    | 0.748370457307   |
| 26 | 0.3381648484657   | 0.73863676355    |
| 28 | 0.314699900401    | 0.729832566394   |

dependence of the ratios:

$$R(\tau,p,N) \equiv \frac{S(\tau,p,N)}{S(0,p,N)} \qquad (3.1)$$

for $\tau = \frac{1}{4}, \frac{1}{3}, \frac{1}{2}, \frac{3}{4}$ fixed and $N = 4, 6..., 28$. The data points for $N \geq 8$ scale in the momentum $p$ except for $p = \pi$. Therefore the resulting curves at fixed $\tau$ represent already the thermodynamical limit $R(\tau, p, N = \infty)$ for $p < \pi$. The curves start with $R = 1$ at $p = 0$, have a minimum at $p = p_0(\tau) \approx \pi/2$ and approach again $R = 1$ for $p = \pi$. The latter is a consequence of the nonintegrable infrared singularity in (1.5). The solid curves in figure 1 represent the prediction of the ansatz (1.6) of Müller *et al.* [15]. This prediction is in good agreement with our finite system results for small $p$ values. Deviations emerge for larger $p$ values, which will be analyzed in the next section.

For increasing values of $\tau$ finite size effects increase. They become visible in the quantities $\rho(t, p, N)$, which are related to the ratios (3.1) via:

$$R(\tau,p,N) = \frac{1}{1+\rho(t,p,N)}. \qquad (3.2)$$

In terms of the variable:

$$t = \sqrt{\omega_1(p)\tau}\exp[\omega_1(p)\tau] \qquad (3.3)$$

the large $t$ behaviour:

$$\rho(t,p,N=\infty) \stackrel{t\to\infty}{\longrightarrow} t \qquad (3.4)$$

is linear. (3.4) originates from the threshold singularity (1.6) which is projected out in (1.8) in the large $\tau$ limit.

The small $t$ behaviour:

$$\rho(t,p,N=\infty) \stackrel{t\to 0}{\longrightarrow} \frac{\rho_2(p)}{\omega_1(p)}t^2 \qquad (3.5)$$

follows from a Taylor expansion of (1.8) around $\tau = 0$. The coefficient:

$$\rho_2(p) \equiv -\frac{d}{d\tau} R(\tau, p, N = \infty)\bigg|_{\tau=0} = \frac{K^{(1)}(p)}{K^{(0)}(p)} \tag{3.6}$$

is given by the first two moments of the dynamical structure factor:

$$K^{(n)}(p) \equiv \int_0^\infty d\omega \, \omega^n S(\omega, p). \tag{3.7}$$

$K^{(0)}(p)$ is just the static structure factor. It has been determined numerically for $\pi/14 \leq p \leq 13\pi/14$ in reference [17]. The first moment is known analytically [13]:

$$K^{(1)}(p) = (1 - \cos p)\epsilon_0 \tag{3.8}$$

where $\epsilon_0 = (4\ln 2 - 1)/3$ is related to the groundstate energy. The coefficient $\rho_2(p)$ is given in table 3.

**Table 3.** The ratio (3.6) of frequency moments $K^{(1)}/K^{(0)}$ versus $p$.

| $p/\pi$ | 1/14 | 1/13 | 1/12 | 1/11 | 1/10 | 1/9 | 1/8 | 1/7 |
|---|---|---|---|---|---|---|---|---|
| $\rho_2(p)$ | 0.7373 | 0.7919 | 0.8552 | 0.9296 | 1.0180 | 1.1249 | 1.2566 | 1.4226 |
| $p/\pi$ | 1/6 | 1/5 | 1/4 | 2/7 | 1/3 | 2/5 | 3/7 | 1/2 |
| $\rho_2(p)$ | 1.6368 | 1.9228 | 2.3173 | 2.5705 | 2.8662 | 3.1930 | 3.2997 | 3.4741 |
| $p/\pi$ | 4/7 | 3/5 | 2/3 | 5/7 | 3/4 | 4/5 | 5/6 | |
| $\rho_2(p)$ | 3.5133 | 3.4912 | 3.3579 | 3.1933 | 3.0353 | 2.7634 | 2.5496 | |
| $p/\pi$ | 6/7 | 7/8 | 8/9 | 9/10 | 10/11 | 11/12 | 12/13 | 13/14 |
| $\rho_2(p)$ | 2.3805 | 2.2432 | 2.1301 | 2.0353 | 1.9545 | 1.8847 | 1.8237 | 1.7698 |

For all values of $p$ and $N$ we observe a very smooth and almost linear behaviour of $\rho$ in $t$, except near $t = 0$, where the quadratic behaviour (3.5) is visible. Finite size effects can be described with high accuracy by:

$$\rho(t, p, N) = \rho(t, p, N = \infty) + \frac{c(t, p)}{N^2} \tag{3.9}$$

as is demonstrated in figure 2(a) for $p/\pi = \frac{1}{3}, \frac{1}{2}, \frac{2}{3}$ and $t = 1, 5, 10$. The coefficient $c(t, p)$ increases with $t$ and with $p$ for $p > \pi/3$, as can be seen in figure 2(b). The resulting estimates $\rho(t, p) \equiv \rho(t, p, N = \infty)$ for the thermodynamical limit are plotted in figure 3 for $p/\pi = \frac{1}{7}, \frac{1}{4}, \frac{2}{7}, \frac{1}{3}, \frac{3}{7}, \frac{1}{2}, \frac{4}{7}, \frac{2}{3}, \frac{5}{7}, \frac{3}{4}, \frac{6}{7}$. Here we have included momentum values $p = k\pi/7$, $k = 1, 2, ..., 6$, which are realized on two systems ($N = 14, 28$) only – assuming that the finite size behaviour is of the form (3.9). Note that $\rho(t, p)$ is strictly monotonic in $p$ and $t$. In order to reach this property we have changed the definition of the variable $t$. (3.3) differs from the definiton given in reference [15] by the factor $\sqrt{\omega_1(p)}$.

The error in our estimate of $\rho(t,p)$ is presumably less than the width of the lines in figure 3. This optimistic view appears to be justified by the clean signal for the finite size behaviour (3.9) seen in the data.

## 4. A modified ansatz for the dynamical structure factors

In figure 4(a) we compare our estimate for $\rho(t,p)$ with the prediction $\rho_M(t,p)$, which follows from the ansatz (1.6). The difference:

$$\Delta_M(t,p) = \rho_M(t,p) - \rho(t,p) \tag{4.1}$$

is negative for $p < \pi/3$ and positive for $p > \pi/3$. It increases with $t$ and $p$ for $p > \pi/3$. The increase of the deviations with $p$ has been observed also in the static structure factors [17]. Before presenting modifications of the ansatz, which are in better agreement with the estimate of the thermodynamical limit let us briefly review the properties of the frequency moments $K^{(n)}(p)$ – defined in (3.7):

(i) The moment (3.8) with $n=1$ is determined by the groundstate energy.

(ii) The higher moments with $n$ odd are known [13] to be polynomials of order $n$ in $\cos p$.

(iii) The moment with $n = -1$ is related to the static susceptibility $\chi(p)$, which yields in the zero momentum limit [12]:

$$K^{(-1)}(p) \xrightarrow{p \to 0} \frac{\chi(0)}{2} = \frac{1}{4\pi^2}. \tag{4.2}$$

(iv) The moment with $n = 0$ is identical with the static structure factor, which has been determined in reference [17] from a finite size analysis in the momentum interval $\pi/14 \leq p \leq 13\pi/14$.

The ansatz of Müller *et al.* [12] respects the first two properties. However, the first and third one cannot be satisfied simultaneously with this ansatz. We therefore would like to propose a modification here:

$$S(\omega,p) = \frac{A(p)}{\sqrt{\omega^2 - \omega_1^2}} \Theta(\omega - \omega_1) \Theta(\hat{\omega}_2 - \omega) \tag{4.3}$$

with a momentum dependent amplitude $A(p)$ and a modified high-frequency cut-off:

$$\hat{\omega}_2(p) \equiv u(p)\omega_2(p). \tag{4.4}$$

Properties (i) and (iii) are satisfied now if we choose:

$$u(p) \xrightarrow{p \to 0} 1.274..., \quad A(p) \xrightarrow{p \to 0} 0.7475... \cdot p. \tag{4.5}$$

This yields a prediction for the static structure factor in the low-momentum limit:

$$K^{(0)}(p) \xrightarrow{p \to 0} 1.084... \cdot \frac{p}{4\pi} \tag{4.6}$$

which is in good agreement with the results of the finite size analysis [17]. For $\pi/14 \leq p \leq 13\pi/14$, we fix $u(p)$ in such a way that the ratio (3.6) of the moments with $n = 1$ and $n = 0$ is reproduced correctly. $u(p)$ is shown in figure 5(a). It turns out to be smaller than 1 for $p > \pi/3$ and larger than 1 for $p < \pi/3$. The extrapolation to $p = 0$ seems to meet the value given in (4.5). Finally, we determine the $p$-dependence of the amplitude $A(p)$ from the moment sum rule (3.8) with $n = 1$. $A(p)$ is shown in figure 5(b). The extrapolation to $p = 0$ is compatible with the behaviour given in (4.5). The values of $A(p)$ and $u(p)$ at $p = \pi$:

$$A(\pi) = 1.88(8), \qquad u(\pi) = 0.63(5), \tag{4.7}$$

have been obtained in reference [17] from a fit to the static structure factor of the form:

$$K^{(0)}(p) \xrightarrow{p \to \pi} -\frac{A(\pi)}{2\pi} \ln\left(1 - \frac{p}{\pi}\right) + const. \tag{4.8}$$

In figure 4(b) we compare the estimate for $\rho(t,p)$ with the prediction $\rho_{MM}(t,p)$, which follows from the ansatz (4.3) with the momentum dependent amplitude $A(p)$ and the modified high-frequency cut-off (4.4). The difference:

$$\Delta_{MM}(t,p) = \rho_{MM}(t,p) - \rho(t,p) \tag{4.9}$$

is obviously much smaller than (4.1), which is shown in figure 4(a). On the other hand (4.9) is not yet zero. In particular for large $t$ and $p = 3\pi/4$ we still observe sizeable deviations form our estimate $\rho(t,p)$ for the thermodynamical limit, leaving room for further improvements of the ansatz (4.3).

## 5. Conclusions

The euclidean time representation (1.8) is particularly suited for a study of finite size effects in dynamical structure factors. In this paper we investigated the spin-1/2 antiferromagnetic Heisenberg model (1.1). We found that finite size effects in the euclidean time representation die out with $N^{-2}$ for those momentum values $p$ where there is a gap (1.4) between the groundstate and the first excited state with momentum $p$. We then compared our estimate for the thermodynamical limit with two predictions. Each of them follows from an ansatz of the type (4.3) for the spectral ($\omega$)- representation of the dynamical structure factor:

- The first ansatz – i.e. (4.3) with momentum independent amplitude $A(p) = A$ and $u(p) = 1$ – has been proposed by Müller *et al.* [12]. It takes into account two-spinon contributions only and respects the properties (i) and (ii) on the frequency moments $K^{(n)}(p)$ listed in section 4. Deviations from our estimate for the thermodynamical limit appear for larger momenta and euclidean times $\tau$.

• In the second ansatz (4.3) the momentum dependences of the amplitude $A(p)$ and of the modification $u(p)$ for the high-frequency cut-off (4.4) have been chosen such that the spectral moments $K^{(n)}(p)$ with $n = -1, 0, 1$ are reproduced. In particular, the zero momentum behaviour of $K^{(-1)}(p)$ and $K^{(1)}(p)$ enforces the vanishing of the amplitude $A(p = 0)$ and a well defined modification (4.5) of the high-frequency cut-off (4.4). The second ansatz is in better agreement with our estimate for the thermodynamical limit of the dynamical structure factor.

We are aware of the fact, that the relevant excitation spectrum in the spin-1/2 antiferromagnetic Heisenberg model is unbounded. One might ask how to interpret the 'effective' high-frequency cut-off (4.4) and the fact that it is above the two-spinon cut-off for $p < \pi/3$ and below for $p > \pi/3$? Adding in the dynamical structure factors high-frequency excitations ($\omega > \omega_2(p)$) – which are not of the two-spinon type – will always lead to an increase of the effective high-frequency cut-off (4.4). Therefore, such contributions are responsible for the increase of the effective high-frequency cut-off (4.4) for $p < \pi/3$. On the other hand they cannot explain the lowering of the effective high-frequency cut-off for $p > \pi/3$.

### Acknowledgments

We thank Prof. Gerhard Müller from the University of Rhode Island for helpful comments on this paper.

A. Fledderjohann was supported by the Deutsche Forschungsgemeinschaft under contract number Mü 810/4-1.

12

**Figure captions**

**Figure 1** The momentum dependence of the ratios $R(\tau, p, N)$ (3.1) at $\tau = \frac{1}{4}, \frac{1}{3}, \frac{1}{2}, \frac{3}{4}$ and $N = 4, 6, ..., 28$. The solid curve is the prediction of reference [12].

**Figure 2** The finite size behaviour (3.9) of $\rho(t, p, N)$ – defined in (3.2): (a) The $N^{-2}$ dependence at $p/\pi = \frac{1}{3}, \frac{1}{2}, \frac{2}{3}$ and $t = 1, 5, 10$. (b) The coefficient $c(t, p)$ versus $t$ (3.3) at fixed $p/\pi = \frac{1}{4}, \frac{1}{3}, \frac{1}{2}, \frac{2}{3}, \frac{3}{4}$.

**Figure 3** The estimate $\rho(t, p)$ of the thermodynamical limit versus $t$ at fixed momenta $p/\pi = \frac{1}{7}, \frac{1}{4}, \frac{2}{7}, \frac{1}{3}, \frac{3}{7}, \frac{1}{2}, \frac{4}{7}, \frac{2}{3}, \frac{5}{7}, \frac{3}{4}, \frac{6}{7}$. The inset shows the magnification for $0 \leq t \leq 1$.

**Figure 4** Comparison of the estimate $\rho(t, p)$ with: (a) the ansatz of Müller *et al.* [12] (b) the modified ansatz (4.3).

**Figure 5** The momentum dependence of the parameters in the modified ansatz (4.3): (a) the modification $u(p)$ of the high-frequency cut-off (4.4) (b) The amplitude $A(p)$.

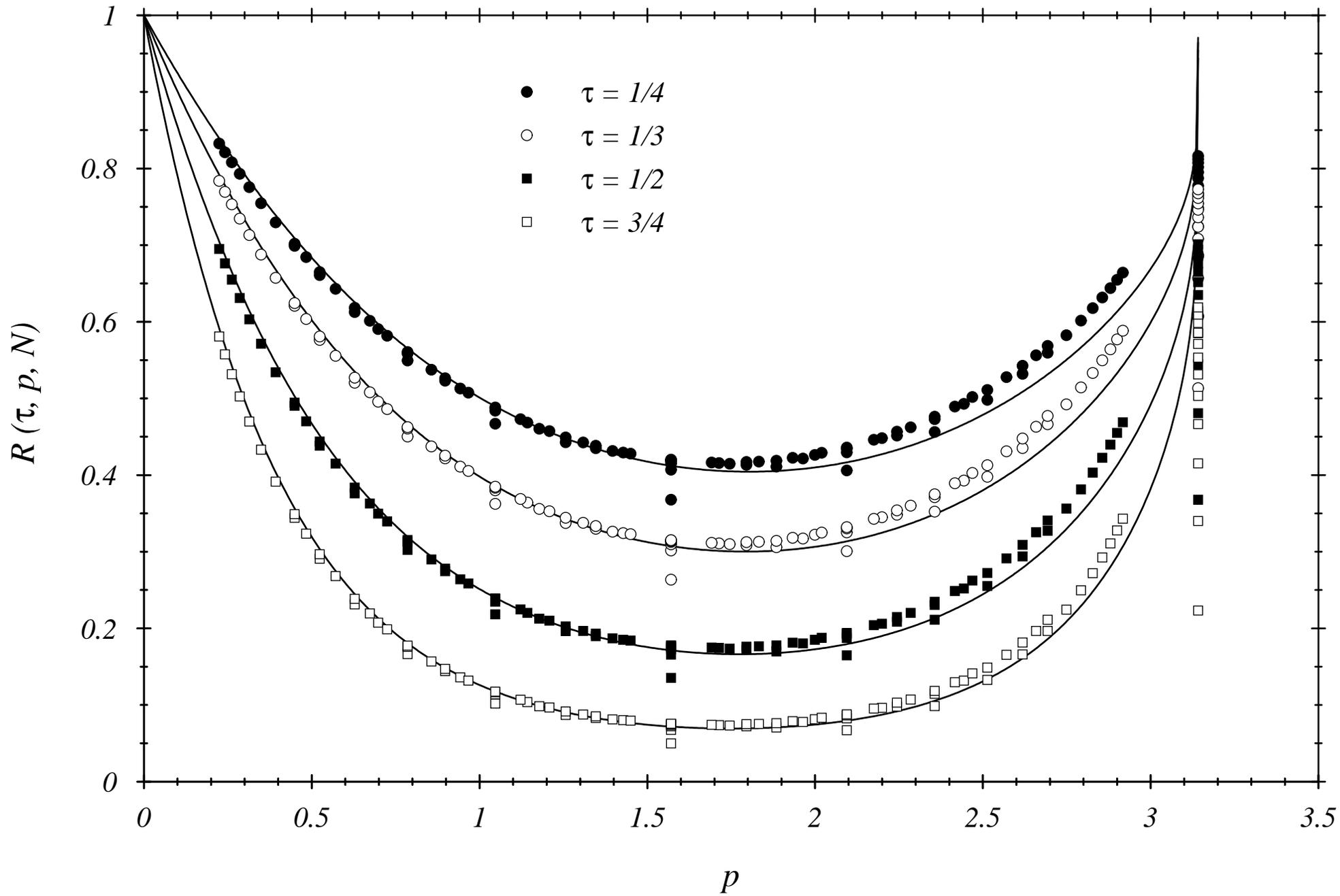

FIG.1

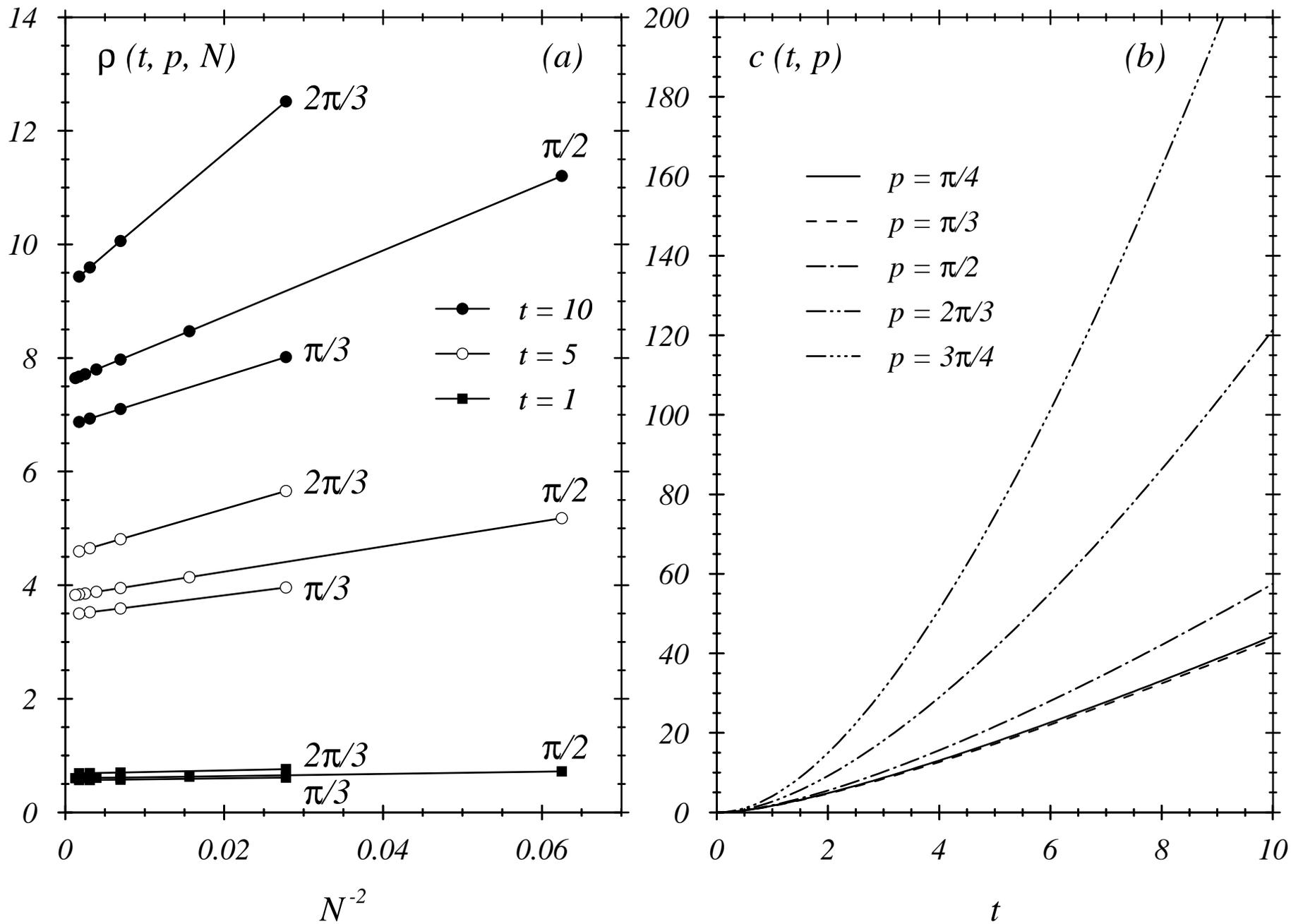

FIG.2

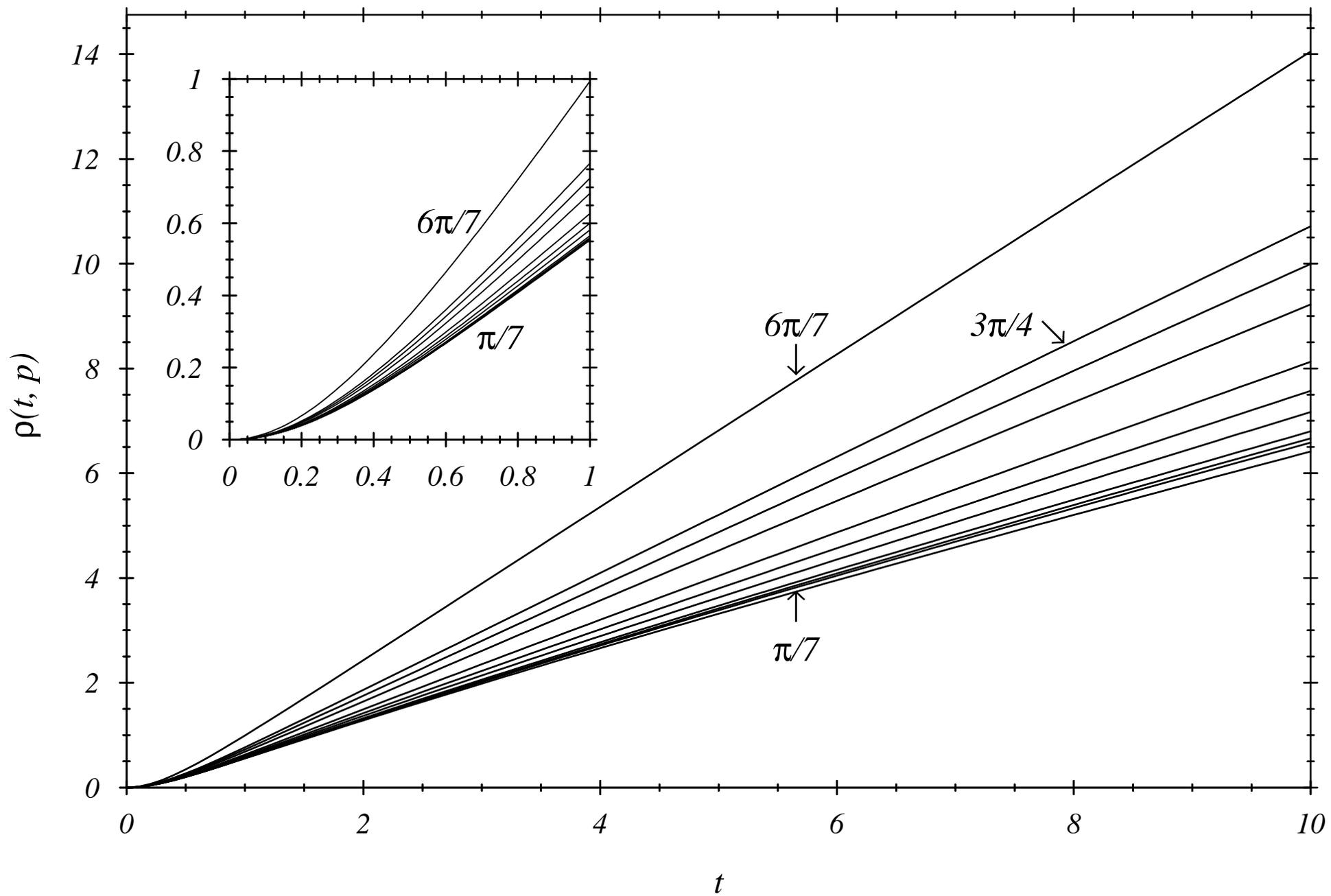

FIG.3

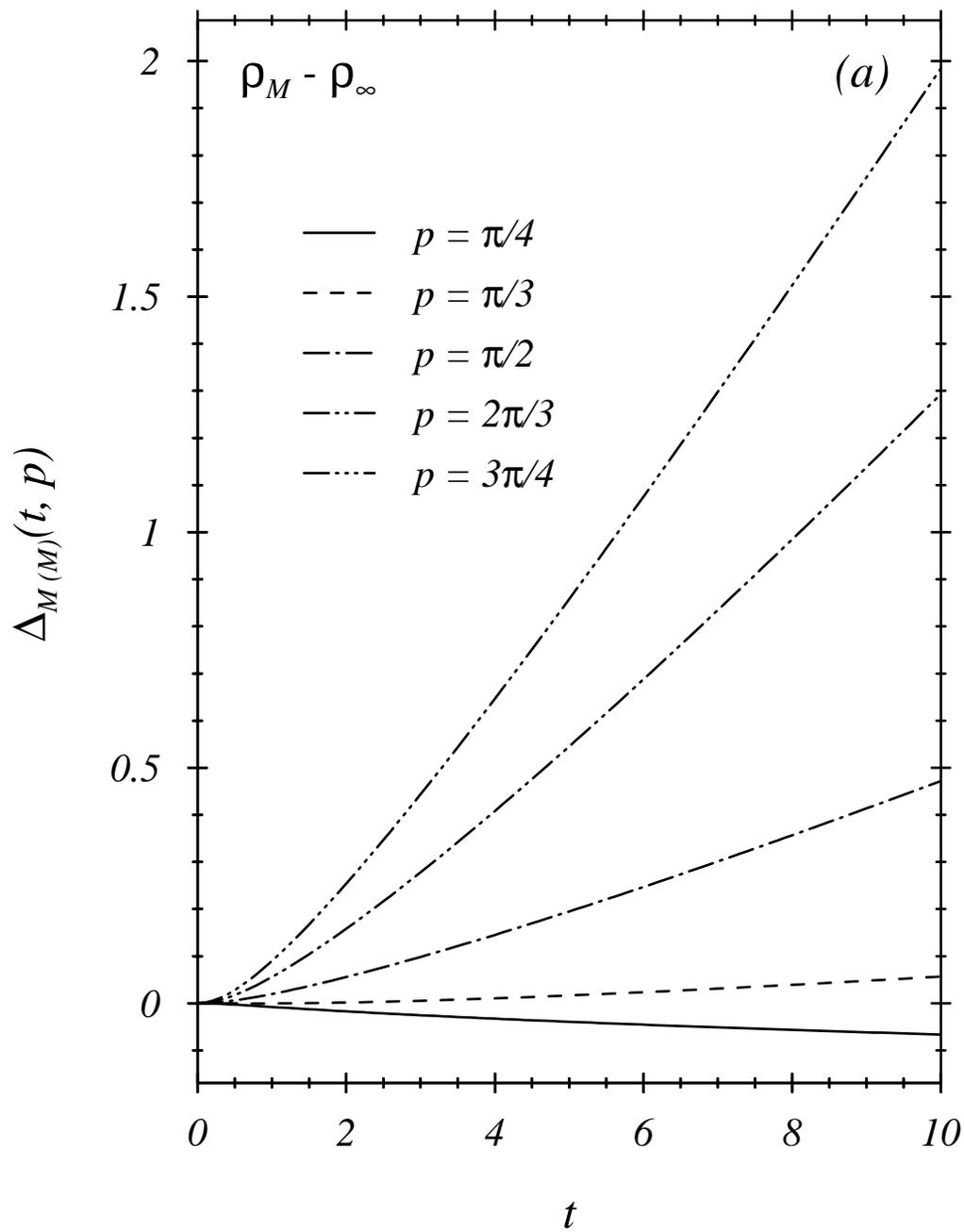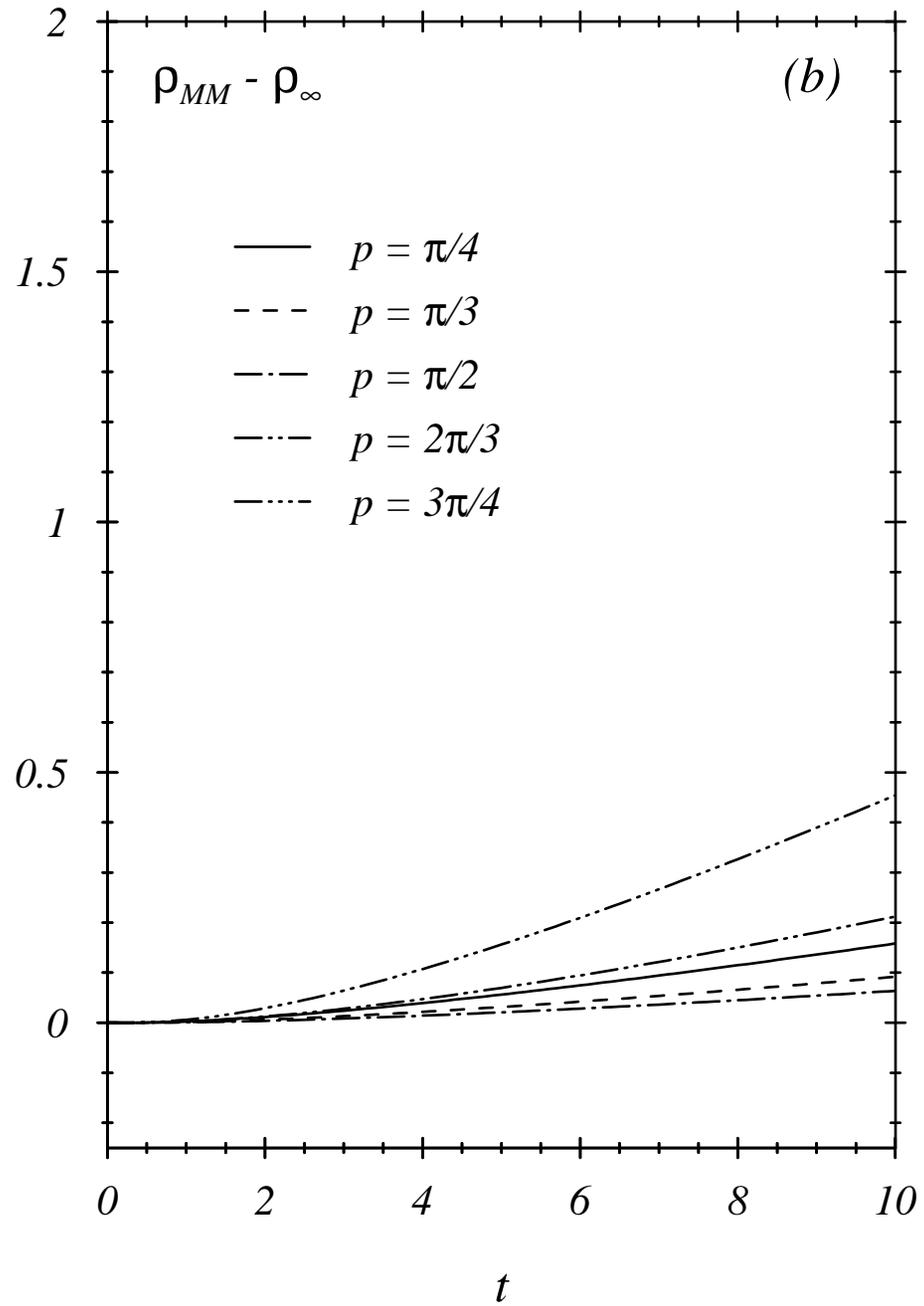

FIG.4

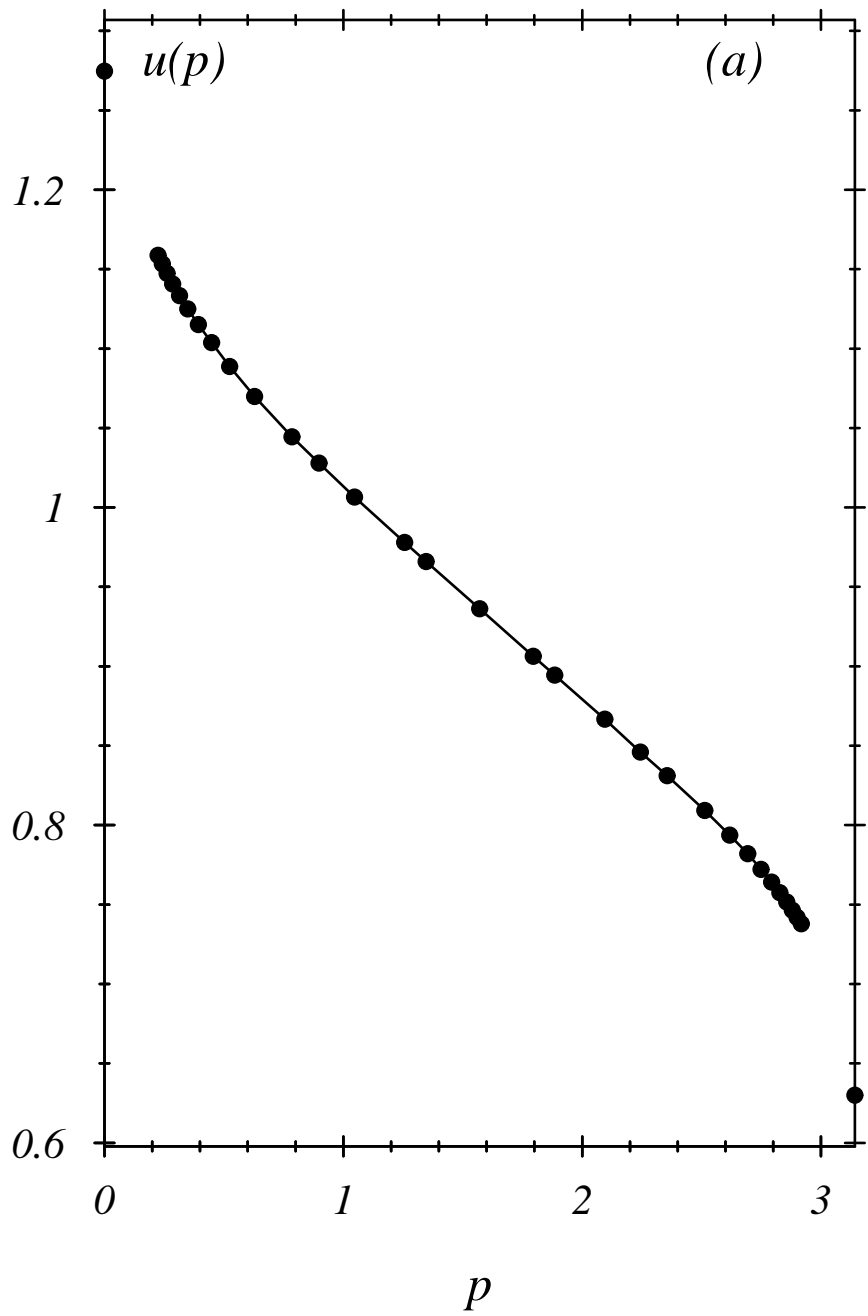 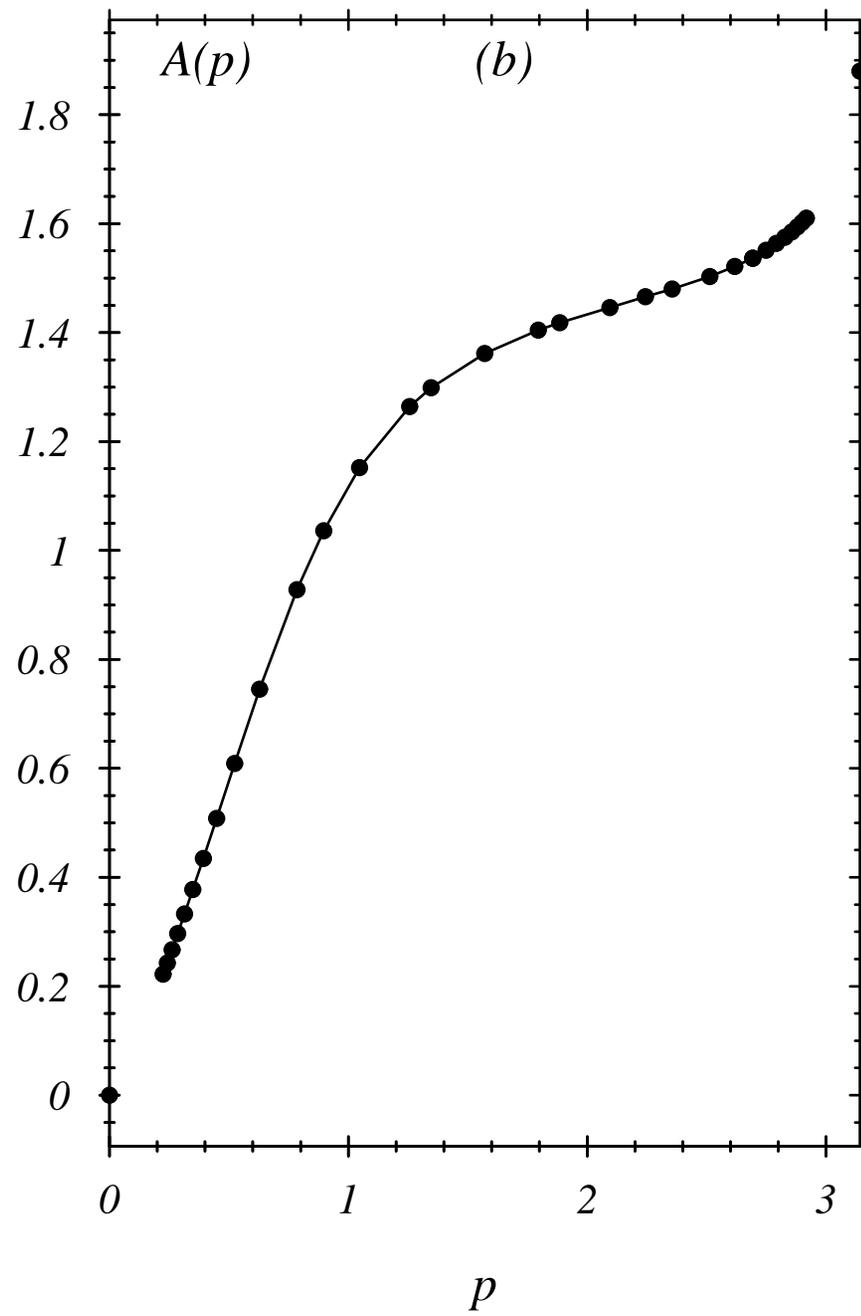

FIG. 5